\documentclass[9pt,cites,graphicx]{article}

\usepackage{epsfig}

\title{Precision molecular spectroscopy for ground state transfer of molecular quantum gases}

\author{Johann G. Danzl,$^a$ Manfred J. Mark,$^a$ Elmar Haller,$^a$ Mattias Gustavsson,$^a$ \\[3mm]
Nadia Bouloufa,$^b$ Olivier Dulieu,$^b$ Helmut Ritsch,$^c$ Russell Hart,$^a$\\[3mm]
and Hanns-Christoph N\"agerl$^a$ \\[3mm]
$^a$ Institut f{\"u}r Experimentalphysik und Zentrum f\"{u}r Quantenphysik, \\
Universit{\"a}t Innsbruck, Technikerstra{\ss}e 25, A--6020 Innsbruck, Austria. \\
E-mail: johann.danzl@uibk.ac.at\\[1mm]
$^b$ Laboratoire Aim\'e Cotton, CNRS, Universit\'e Paris-Sud, B\^{a}t. 505, \\
91405 Orsay Cedex, France. \\[1mm]
$^c$ Institut f\"ur Theoretische Physik und Zentrum f\"{u}r Quantenphysik, \\
Universit{\"a}t Innsbruck, Technikerstra{\ss}e 25, A--6020 Innsbruck, Austria.}

\begin{document}
\maketitle
\renewcommand{\thefootnote}{\fnsymbol{footnote}}

\noindent One possibility for the creation of ultracold, high-phase-space-density quantum gases of mole\-cules in the rovibrational ground state relies on first associating weakly-bound molecules from quantum-degene\-rate atomic gases on a Feshbach resonance and then transfering the molecules via several steps of coherent two-photon stimulated Raman adiabatic passage (STIRAP) into the rovibronic ground state. Here, in ultracold samples of Cs$_2$ Feshbach molecules produced out of ultracold samples of Cs atoms, we observe several optical transitions to deeply bound rovibrational levels of the excited $0_u^+$ molecular potentials with high resolution. At least one of these transitions, although rather weak, allows efficient STIRAP transfer into the deeply bound vibrational level $|v\!=\!73>$ of the singlet X$^1\Sigma_g^+$ ground state potential, as recently demonstrated \cite{Danzl2008}. From this level, the rovibrational ground state level $|v\!=\!0, J\!=\!0\!>$ can be reached with one more transfer step. In total, our results show that coherent ground state transfer for Cs$_2$ is possible using a maximum of two successive two-photon processes or one single four-photon STIRAP process.

\section{Introduction}
\label{intro}
Ultracold and dense molecular samples in specific deeply bound rovibrational levels are of high interest for fundamental studies in physics and chemistry. They are expected to find applications in high resolution spectroscopy and fundamental tests \cite{Zelevinsky2008,DeMille2008}, few-body collisional physics \cite{Staanum2006,Zahzam2006}, ultracold chemistry \cite{Krems2005}, quantum processing \cite{DeMille2002}, and in the field of dipolar quantum gases and dipolar Bose-Einstein condensation \cite{Goral2002,Baranov2002}. Ideally, full control over the molecular wave function is desired, i.e. full (quantum) control over the internal and external degrees of freedom of the molecules. High phase space densities are needed for molecular quantum gas studies. For many of the envisaged studies and applications, initial preparation of the molecular sample in the rovibronic ground state, i.e. the lowest energy level of the electronic ground state, is desired. Only in this state one can expect sufficient collisional stability.

But how is it possible to produce dense samples of ultracold molecules in the rovibrational ground state? Laser cooling of atoms, which has lead to the production of quantum degenerate atomic Bose and Fermi gases \cite{Southwell2002}, can so far not be adapted to the case of molecular systems as suitable cycling transitions are not available. Versatile non-optical cooling and slowing techniques such as buffer gas cooling and Zeeman slowing in combination with molecule trapping \cite{Doyle2004,Krems2008,Meerakker2008} have been developed, but high molecular densities and in particular high phase space densities are yet to be reached. An alternative route to producing ultracold molecular samples is given by first producing ultracold atomic samples and then associating molecules out of the atomic sample. While this technique is so far limited to the production of selected species of dimer molecules, it has the advantage that ultra-low temperatures and high particle densities are easily inherited from the atomic precursor sample. There are essentially two association techniques, photoassociation \cite{Jones2006} and magnetically induced Feshbach association \cite{Koehler2006,Ferlaino2008}. In photoassociation experiments \cite{Nikolov2000,Sage2005,Viteau2008,Deiglmayr2008}, ultracold samples of deeply bound molecules have been created. Additional techniques such as vibrational cooling \cite{Viteau2008} should allow selective pumping into the rovibrational ground state and open up the prospect for high molecular phase space densities. In Feshbach association experiments \cite{Regal2003,Herbig2003}, high-density samples of weakly bound molecules are produced. For dimer molecules composed of Fer\-mions, collisional stability of the highly excited molecules is assured as a result of a Pauli blocking effect, and molecular Bose-Einstein condensation could be achieved in the limit of extremely weak binding \cite{Fermi2008}.

Here, we are interested in combining the techniques of Feshbach association and coherent molecular state transfer to produce quantum gases of molecules in the rovibrational ground state $|v\!=\!0, J\!=\!0\!>$ of the lowest electronic state. As usual, $v$ and $ J $ are the vibrational and rotational quantum numbers, respectively. The molecules, produced on a Feshbach resonance and hence initially very loosely bound, are to be transferred in a few successive steps of coherent two-photon laser transfer to the rovibrational ground state, acquiring more and more binding energy in each step. The general idea is sketched in Fig.1A for the case of Cs$_2$. Each two-photon step involves an excited state level. Population transfer into this level needs to be avoided to prevent loss due to spontaneous emission. One possibility is to use the technique of stimulated Raman adiabatic passage (STIRAP) \cite{Bergmann1998}, which is very robust and largely insensitive to laser intensity fluctuations. The scheme has several advantages. First, production of Feshbach molecules out of a quantum degenerate atomic sample can be very efficient \cite{Mark2005}.
Second, the optical transition rate on the first transition starting from the Feshbach molecules is greatly enhanced in comparison to the free atom case. Further, the scheme is fully coherent, not relying on spontaneous processes, allowing high state selectivity, and involving only a comparatively small number of intermediate levels. A ground state binding energy of typically 0.5 eV for an alkali dimer can be removed essentially without heating the molecular sample, as the differential photon recoil using pairwise co-propagating laser beams driving the two-photon transitions is very small. If losses and off-resonant excitations can be avoided, the scheme essentially preserves phase space density and coherence of the initial particle wave function, allowing the molecular sample to inherit the high initial phase space density from the atomic precursor sample.

Certainly, several challenges have to be met: Going from weakly bound Feshbach to tightly bound ground state molecules corresponds to a large reduction in internuclear distance. Consequently, the radial wave function overlap between successive levels is small, and a compromise has to be found between the number of transitions and the minimum tolerable wave function overlap. To keep the complexity of the scheme low, one or at most two two-photon transitions are desirable. Accordingly, suitable intermediate levels have to be identified that allow a balanced division of wave function overlap, as given by the Franck-Condon factors, between the different transitions. For example, for a four-photon transition scheme with Cs$_2$ as shown in Fig.1A the Franck-Condon factors are all on the order of $10^{-6}$. We emphasize that the identification of the first excited level and hence of the first transition starting from the Feshbach molecules is of crucial importance. Detailed calculations determining the wave function overlap are generally missing, and estimates on the Franck-Condon factors using hypothetical last bound states of either the singlet or triplet potentials of an alkali dimer molecule do not necessarily reflect the transition dipole moments adequately. In addition, for electronic molecular states or energy regions where spectroscopic data is missing, the precise energy of the excited state levels above the atomic threshold is known only with a large uncertainty which can approach the vibrational spacing of a few nanometers. Hence, considerable time has to be spent on searching for weak transitions starting from the initial Feshbach molecules.

In a pioneering experiment, Winkler {\it et al.} \cite{Winkler2007} demonstrated that the STIRAP technique can efficiently be implemented with quantum gases of weakly bound Feshbach molecules. In this work, the transferred molecules, in this case Rb$_2$, were still weakly bound with a binding energy of less than $10^{-4}$ of the binding energy of the rovibronic ground state, and the intermediate excited state level was close to the excited-atom asymptote. Here, we observe several optical transitions starting from a weakly bound Feshbach level to deeply bound rovibrational levels of the mixed excited $($A$^1\Sigma_u^+ - $b$^3\Pi_u) \ 0_u^+$ molecular potentials of the Cs$_2$ molecule in a wavelength range from 1118 to 1134 nm, far to the red of the atomic D$_1$ and D$_2$ transitions. The Cs$_2$ molecular potentials are shown in Fig.1A. We observe the levels as loss from an ultracold sample of Cs$_2$ Feshbach molecules as shown in Fig.1B. We observe two progressions, one that we attribute to the $($A$^1\Sigma_u^+ - $b$^3\Pi_u) \ 0_u^+$ potentials and one that we associate to the triplet $(1)^3\Sigma_g^+$ potential. From the loss measurements, we determine the transition strengths and find that the stronger transitions should be suitable for STIRAP to an intermediate, deeply bound rovibrational level of the singlet X$^1\Sigma_g^+$ potential with $v\!=\!73$. Recently, we could implement STIRAP into $|v\!=\!73, J\!=\!2\!>$ \cite{Danzl2008}. For the case of the dimer molecule KRb, Ni {\it et al.} \cite{Ni2008} could demonstrate quantum gas transfer all the way into the rovibrational ground state $|v\!=\!0, J\!=\!0>$ of the singlet X$^1\Sigma^+$ molecular potential. Here, the transfer could be achieved in only a single step as a result of the favorable run of the excited state potentials, which is generally the case for heteronuclear molecules composed of alkali atoms \cite{Stwalley2004}. Also recently, transfer to the rovibrational ground state of the lowest triplet state a$^3\Sigma_u^+$ of Rb$_2$ could be achieved \cite{Lang2008}.

\section{Preparation of a sample of weakly bound Feshbach molecules}
\label{exper1}
We produce ultracold samples of molecules on two different Feshbach resonances, one near 1.98 mT and one near 4.79 mT \cite{Mark2007}. In both cases, essentially following the procedure detailed in Ref.\cite{Weber2003}, we first produce an ultracold sample of typically $2\times10^5 $ Cs atoms in the lowest hyperfine sublevel $F\!=\!3, \ m_F\!=\!3$ in a crossed optical dipole trap. As usual, $F$ is the atomic angular momentum quantum number, and $m_F$ its projection on the magnetic field axis. The trapping light at 1064.5 nm is derived from a single-frequency, highly-stable Nd:YAG laser. The offset magnetic field value for evaporative cooling is 2.1 mT. We support optical trapping by magnetic levitation with a magnetic field gradient of 3.1 mT/cm.
We then produce weakly bound Feshbach molecules out of the atomic sample \cite{Herbig2003}.
We produce a sample every 8 s, i.e. our spectroscopic measurements are performed at a rate of one data point every 8 s.
In order to be able to search for optical transitions over large frequency ranges it is advantageous to work with the shortest possible sample preparation times. For this reason we stop evaporative cooling slightly before the onset of Bose-Einstein condensation (BEC), which also makes sample preparation somewhat less critical. The temperature of the initial atomic sample is then typically about 100 nK. At higher temperatures and hence lower phase space densities the molecule production efficiency is reduced, so that there is a trade off between ease of operation and molecule number. We note that for our ground state transfer experiments reported in Ref.\cite{Danzl2008} we produce a pure atomic BEC at the expense of longer sample preparation times.

The spectrum of weakly-bound Feshbach levels near the two-free-atom asymptote is shown in Fig.2 \cite{Mark2007}. For molecule production at the Feshbach resonance at 4.79 mT, we first ramp the magnetic field from the BEC production value to 4.9 mT, about 0.1 mT above the Feshbach resonance. We produce the molecular sample on a downward sweep at a typical sweep rate of 0.025 mT/ms. The resulting ultracold sample contains up to 11000 molecules, immersed in the bath of the remaining ultracold atoms. The resonance at 4.79 mT is a $d$-wave resonance \cite{Mark2007}, and hence the molecules are initially of $d$-wave character, i.e. $\ell \! = \! 2$, where $\ell$ is the quantum number associated with the mechanical rotation of the nuclei. However, there is a weakly bound $s$-wave Feshbach state ($|s> = \! |\ell \! = \! 0>$) belonging to the open scattering channel right below threshold. This state couples quite strongly to the initial $d$-wave state, resulting in an avoided state crossing (as shown in the inset to Fig.2), on which the molecules are transferred to the $s$-wave state $|s\!>$ upon lowering the magnetic field \cite{Mark2007,Danzl2008}. Upon further lowering the magnetic field to less than 2.0 mT, the molecules acquire more and more character of a closed channel $s$-wave state on a second, very broad avoided crossing. Here, we perform spectroscopy in this transition range from open channel to closed channel $s$-wave character. At a magnetic field value of 2.0 mT, the binding energy of the molecules is near $ 5 $ MHz$\times h$ with respect to the $F\!=\!3, m_F\!=\!3$ two-atom asymptote, where $h$ is Planck's constant.

For molecule production at the Feshbach resonance at 1.98 mT, we simply ramp the magnetic field down from the initial BEC production value. Again, we produce an ultracold molecular sample with about 11000 molecules. The molecules in $|g\!>$ have $g$-wave character, i.e. $\ell \! = \! 4$. When we lower the magnetic field to 1.6 mT, the binding energy of the molecules is also near $ 5 $ MHz$\times h$ with respect to the $F\!=\!3, m_F\!=\!3$ two-atom asymptote.

For spectroscopy, we release the molecules from the trap after magnetic field ramping is completed and perform all subsequent experiments in free flight without any other light fields on except for the spectroscopy laser.

For molecule detection in both cases, we reverse the magnetic field ramps \cite{Herbig2003}. The $g$-wave molecules are dissociated on the $g$-wave Feshbach resonance at 1.98 mT, and the $s$-wave molecules are dissociated on the $d$-wave Feshbach resonance at 4.79 mT. Prior to the reverse magnetic field ramp, we apply a magnetic field gradient of 3.1 mT/cm for about 5 ms to separate the molecular from the atomic sample in a Stern-Gerlach-type experiment. Finally, we detect atoms by standard absorption imaging. The minimum number of molecules that we can detect is on the order of 200 molecules.

\section{Spectroscopy}
\label{exper2}

We perform optical spectroscopy on Feshbach molecules in the wavelength region around 1125 nm. Based on symmetry considerations, there are two sets of electronically excited states that we address in the spectroscopic measurements presented here, namely the $($A$^1\Sigma_u^+ - $b$^3\Pi_u) \ 0_u^+$ coupled state system and the $(1)^3\Sigma_g^+$ electronically excited states. We first discuss transitions to the $0_u^+$ coupled state system. Transitions to the latter state are discussed in Sec. \ref{exper:triplet}.

\subsection{Transitions to the $($A$^1\Sigma_u^+ - $b$^3\Pi_u) \ 0_u^+$ coupled electronically excited states}
\label{exper:0u+}

We are primarily interested in transitions from Feshbach levels to rovibrational levels of the $($A$^1\Sigma_u^+ - $b$^3\Pi_u) \ 0_u^+$ electronically excited states. In the heavy alkali dimers, most notably in Cs$_2$, the A$^1 \Sigma_u^+$ state and the b$^3 \Pi_u$ state are strongly coupled by resonant spin-orbit interaction \cite{Dulieu1995,Amiot1999}, yielding the $0_u^+$ coupled states in Hund's case (c) notation. The singlet component of the $0_u^+$ states allows us to efficiently couple to deeply bound X$^1\Sigma_g^+$ state levels, specifically to the $|v\!=\!73, J\!=\!2\!>$ level of the ground state potential, as has recently been shown in a coherent transfer experiment \cite{Danzl2008}. We have chosen to do spectroscopy in the wavelength range of 1118 nm to 1134 nm above the 6S$_\frac{1}{2}$+6S$_\frac{1}{2}$ dissociation threshold of the Cs$_2$ dimer. This corresponds to a detuning of roughly 2300 cm$^{-1}$ from the cesium D$_1$ line and to an energy range of approximately 12572 cm$^{-1}$ to 12450 cm$^{-1}$ above the rovibronic ground state X$^1\Sigma_g^+$ $|v\!=\!0, J\!=\!0\!>$. This region was chosen in order to give a balanced distribution of transition dipole moments in a 4-photon transfer scheme to the rovibronic ground state. In addition, the wavelengths of the four lasers used in the transfer experiments were chosen such that they lie within the energy range covered by the infrared fiber-based frequency comb that we use as a frequency reference in the state transfer experiments.

The transitions of interest here lie outside the energy regions for which Fourier transform spectroscopic
data was obtained at Laboratoire Aim\'e Cotton from transitions to the X$^1\Sigma_g^+$ state \cite{Salami2008}. The vibrational progression of the $0_u^+$ states is highly perturbed by the resonant spin-orbit coupling and exhibits an irregular vibrational spacing. Molecular structure calculations are complicated by the spin-orbit coupling and calculated term values are highly sensitive to the coupling.
Prior to the experiments discussed here the absolute energies of the vibrational levels of the $($A$^1\Sigma_u^+ - $b$^3\Pi_u) \ 0_u^+$ excited state levels were poorly known in the region of interest from 1118 nm to 1134 nm. We therefore perform a broad range search by irradiating the weakly-bound Feshbach molecules at a fixed wavelength for a certain irradiation time $\tau$ of up to $\tau \! = \! 6$ ms and by recording the number of remaining molecules as a function of laser frequency. In one run of the experiment one particular laser frequency is queried. We thus take data points at the repetition rate of our experiment, which is given by the sample preparation time of 8 seconds.
Based on the available laser intensity from $L_1$ and an estimate of the dipole transition moments for the strongest expected lines, we chose a frequency step size of about 100 MHz to 150 MHz for initial line searching.
We obtain the laser light at 1118 nm - 1134 nm from a grating-stabilized external cavity diode laser. For coarse frequency scanning, the laser is free running and tuned via a piezoelectric element on the grating of the laser. For more precise measurements, we lock the laser to a narrow-band optical resonator that can be tuned via a piezoelectric element.
Fig.3 A shows a typical loss spectrum starting from Feshbach state $|s\!\!>$ for excitation near 1126 nm, measured at a magnetic field of 1.98 mT.  In this particular case we find three resonances, which we associate with the rotational splitting of the excited state level, $J=5,3,1$, where $J$ is the rotational quantum number. Based on molecular structure calculations we identify this level as the 225th one of the $0_u^+$ progression with an uncertainty of about two in the absolute numbering. We zoom in on these three transitions in Fig.3 B, C, and D and record loss resonances at reduced laser intensity in order to avoid saturation of the lines. For these measurements, the laser is locked to the narrow-band optical resonator and the resonator in turn is stabilized to the optical frequency comb to assure reproducibility and long term frequency stability. As one can expect, the loss is strongest on the transition to the $|J\!=\!1>$ level, and it is weakest on the transition to $|J\!=\!5>$. All lines have an excited state spontaneous decay rate of around $2 \pi \times 2$ MHz, in agreement with the typical expected lifetimes of excited molecular levels. The transition to $|J\!=\!1>$ shown in Fig.3 D is of special interest to the current work. It has been used as intermediate excited state level for coherent transfer to X$^1\Sigma_g^+$ $|v\!=\!73, J\!=\!2\!>$ in our recent experiments \cite{Danzl2008}.

By fitting a two level model that takes into account decay from the upper level to a series of such measurements obtained with different laser intensities, we determine the transition strength as given by the normalized Rabi frequency. 
As the Feshbach molecules scatter photons and spontaneously decay to other molecular levels, the number of Feshbach molecules $N$ decays as a function of laser detuning $\Delta_1$ according to $N(\Delta_1)=N_0 \exp{(-\tau \Omega_1^2/(\Gamma(1+4\pi^2\Delta_1^2/\Gamma^2)))}$, where $N_0$ is the molecule number without laser irradiation and $\tau$ is the irradiation time. From the fit we obtain the Rabi frequency on resonance $\Omega_1$ and the excited state spontaneous decay rate $\Gamma$.
We determine the normalized Rabi frequency to $\Omega_1\!=\!2\pi\!\times\!2$ kHz $ \sqrt{I/(\mathrm{mW/cm}^2)}$ for $|J\!=\!1>$, where $I$ is the laser intensity.  This value is sufficient to perform STIRAP given the available laser power \cite{Danzl2008}. The corresponding transition strengths for $|J\!=\!3>$ and $|J\!=\!5>$ are $\Omega_1\!=\!2\pi\!\times\!0.3$ kHz $ \sqrt{I/(\mathrm{mW/cm}^2)}$ and $\Omega_1\!=\!2\pi\!\times\!0.1$ kHz $ \sqrt{I/(\mathrm{mW/cm}^2)}$, respectively. The absolute values of these transition strengths bear an estimated uncertainty of 20 \% because the laser beam parameters for the spectroscopy laser are not well determined.

We also record the time dependence of the molecular loss on some of the stronger lines. For this, we step the laser irradiation time $\tau$ from $0$ to $150 \ \mu$s, while laser $L_1$ is kept on resonance. The result is shown in Fig.4 A for the transition at 1126.173 nm for two different values of the excitation laser intensity.

We note that the transition strength for a particular line starting from Feshbach level $|s\!\!>$ strongly depends on the value of the magnetic field, as evidenced in Fig.4 B. Loss resonances for the transition at 1126.173 nm at 1.9 mT and 2.2 mT are shown. For ground state transfer \cite{Danzl2008}, we choose a magnetic field of around 1.9 mT, which is somewhat below the magnetic field region where state $|s\!>$ is strongly curved, but above the avoided state crossing with state $|g_2\!>$, as seen in Fig.2. The pronounced bending of $|s\!\!>$ is the result of a strong avoided crossing between two s-wave Feshbach levels \cite{Mark2007}. For magnetic field values beyond 3.0 mT the level $|s\!\!>$ can be associated to the $F_1\!=\!3, F_2\!=\!3$ asymptote, where $F_i, i\!=\!1,2$, is the atomic angular momentum quantum number of the $i$-th atom, respectively. Below 2.0 mT the level $|s\!\!>$ can be associated to the $F_1\!=\!4, F_2\!=\!4$ asymptote. It is hence of closed channel character and much more deeply bound with respect to its potential asymptote, effectively by twice the atomic hyperfine splitting, improving the radial wave function overlap with the excited state levels. This increases the transition strength. {Trivially, the resonance frequency is shifted as the binding energy is reduced for larger magnetic field values}. Coupling to the excited state level is reduced from $\Omega_1\!=\!2\pi\!\times\!2$ kHz $ \sqrt{I/(\mathrm{mW/cm}^2)}$ to $\Omega_1\!=\!2\pi\!\times\!1$ kHz $ \sqrt{I/(\mathrm{mW/cm}^2)}$ when the magnetic field is changed from 1.9 mT to 2.2 mT.

As will be discussed in Sec.\ref{concl} it is advantageous to be able to choose different Fesh\-bach states as a starting state for ground state transfer experiments. Therefore, we probe transitions from Feshbach level $|g\!>$ to $($A$^1\Sigma_u^+ - $b$3\Pi_u) \ 0_u^+$ levels. Fig.5 shows loss resonances to the same excited state levels as shown in Fig.3, only that now the initial Feshbach level is $|g\!\!>$ instead of $|s\!\!>$. In this case, the transition to $|J\!=\!3>$ is the strongest, while the transition to $|J\!=\!1>$ is very weak, but can be detected. A comparison of the transition strengths from $|g\!>$ to the excited state level $|J\!=\!3\!>$, giving $\Omega_1\!=\!2\pi\!\times\!1$ kHz $ \sqrt{I/(\mathrm{mW/cm}^2)}$ versus $|s\!>$ to $|J\!=\!1>$ giving $\Omega_1\!=\!2\pi\!\times\!2$ kHz $ \sqrt{I/(\mathrm{mW/cm}^2)}$ shows that level $|g\!>$ could also be potentially used as a starting level for coherent population transfer to deeply bound levels of the ground state but requires longer STIRAP times in order to assure sufficient adiabaticity \cite{Bergmann1998}. The $|J\!=\!3\!>$ excited state level in turn couples to $|J\!=\!2\!>$ in the ground state, as in previous work \cite{Danzl2008}.

In addition to the transition near 1126 nm we find a series of other excited state levels that we assign to the $($A$^1\Sigma_u^+ - $b$^3\Pi_u) \ 0_u^+$ coupled state system. These are listed in Table \ref{tab1}. The assignment to either the $($A$^1\Sigma_u^+ - $b$^3\Pi_u) \ 0_u^+$ system or to the $(1)^3\Sigma_g^+$ electronically excited state discussed below is primarily based on the spacing between neighboring vibrational levels and in addition on the pattern of loss resonances associated with each particular vibrational level. Resonant spin-orbit coupling in the case of the $0_u^+$ states leads to an irregular vibrational spacing. In contrast, the  $(1)^3\Sigma_g^+$ state is not perturbed by spin-orbit interaction and therefore has a regular vibrational progression.
The levels near 1126 nm and near 1123 nm have been used to detect dark resonances with deeply bound levels of the X$^1\Sigma_g^+$ state \cite{Danzl2008}. The ability to couple to these essentially purely singlet ground state levels unambiguously assigns the corresponding excited state levels to the $0_u^+$ system. The data given in Table \ref{tab1} does not represent a fully exhaustive study of the $($A$^1\Sigma_u^+ - $b$^3\Pi_u) \ 0_u^+$ coupled states in the wavelength range of interest. In fact, for the most part we observe those levels of the $0_u^+$ system that have a dominant A$^1\Sigma_u^+$ state contribution, as determined from molecular structure calculations.

\subsection{Transitions to the $(1)^3\Sigma_g^+$ electronically excited state}
\label{exper:triplet}

The Feshbach levels that serve as starting levels for the spectroscopy are of mixed X$^1\Sigma_g^+$ and a$^3\Sigma_u^+$ character. In the wavelength range explored here, excitation to the  $(1)^3\Sigma_g^+$ electronically excited triplet state is possible from the a$^3\Sigma_u^+$ component of the Feshbach molecules. In fact, for a heavy molecule as Cs$_2$, the  $(1)^3\Sigma_g^+$ state is better described by the two separate electronic states $0_g^-$ and $1_g$, denoted by the Hund's case $(c)$ notation. The $(1)^3\Sigma_g^+$  has been previously studied by Fourier transform spectroscopy \cite{Amiot1985}. This state is not of prime interest for the present work as transitions from this state down to the X$^1\Sigma_g^+$ ground state are expected to be strongly suppressed, but would be important for STIRAP transfer into the rovibrational ground state level of the shallow triplet a$^3\Sigma_u^+$ potential \cite{Lang2008}. Certainly, it is important to be able to distinguish rovibrational levels belonging to the $(1)^3\Sigma_g^+$ state from the ones belonging to the $0_u^+$ system, because otherwise time would be wasted in searching for ground state dark resonances that are very weak or even do not exist. Fig.6 A shows a typical loss spectrum for one of the lines that we detected near 1127.37 nm. Due to hyperfine splitting, levels of triplet character exhibit a much richer substructure than the $0_u^+$ levels used for ground state transfer. Several components can be identified as a result of rotational and excited state hyperfine splitting. Zoomed-in regions are shown in Fig.6 B, C, D, and E. We have observed a regularly spaced series of optical transitions which we attribute to the $(1)^3\Sigma_g^+$ excited state as listed in Table \ref{tab1}. 
The levels are well reproduced by molecular structure calculations using the Dunham coefficients from Ref.\cite{Amiot1985}. The vibrational numbering used here is the same as in that work. However, it relies on the absolute energy position of the potential, T$_e$, which was not determined precisely in Ref. \cite{Amiot1985}.

\section{Conclusion}
\label{concl}
We have performed optical spectroscopy starting from weakly bound Cs$_2$ Feshbach molecules into deeply bound rovibrational levels of the mixed excited state $0_u^+$ system and the excited triplet $(1)^3\Sigma_g^+$ state. At least one of the observed transitions, namely the one at 1126.173 nm starting from the Feshbach level $|s\!\!>$ at an offset magnetic field value of 1.9 mT to the excited level $|v'\!=\!225, J\!=\!1\!>$ of the $0_u^+$ system, is strong enough to allow efficient STIRAP transfer into deeply bound rovibrational levels of the singlet X$^1\Sigma_g^+$ ground state potential. The use of this transition for STIRAP has recently been demonstrated in Ref.\cite{Danzl2008}. In that work, the deeply bound rovibrational level $|v\!=\!73, J\!=\!2\!>$ of the X$^1\Sigma_g^+$ ground state potential was populated in the molecular quantum gas regime with 80\% efficiency. The rovibrational ground state $|v\!=\!0, J\!=\!0\!>$ of the X$^1\Sigma_g^+$ ground state potential can thus be reached from the atomic threshold with a maximum of two two-photon STIRAP transfers. Dark resonances connecting $|v\!=\!73, J\!=\!2\!>$ to $|v\!=\!0, J\!=\!0\!>$ have recently been observed \cite{Mark2008}, and two-step STIRAP into $|v\!=\!0, J\!=\!0>$ has recently been implemented \cite{Danzl2008b}. For future experiments, the use of Feshbach level $|g\!\!>$ as the initial state might be advantageous. Level $|g\!\!>$ can be more easily populated, as the Feshbach resonance connected to this level is at a low magnetic field value of 1.98 mT \cite{Mark2007}, where the atomic background scattering length has a moderate value of 155 a$_0$, where a$_0$ is Bohr's radius. The use of this resonance avoids excitation of collective motion of the atomic BEC as a result of a large mean field interaction near the Feshbach resonance at 4.79 mT \cite{Danzl2008}, where the atomic background scattering length is about 935 a$_0$. The transition starting from level $|g\!\!>$ appears to be strong enough to allow STIRAP, this time via the excited state level $|v'\!=\!225, J\!=\!3\!>$ of the $0_u^+$ system. An attractive strategy for the production of a BEC of ground state molecules relies on the addition of a three-dimensional optical lattice. Starting from the atomic BEC, pairs of atoms at individual lattice sites can be produced in a superfluid-to-Mott-insulator transition \cite{Greiner2002} with high efficiencies of up to 50\% \cite{Duerr2008}. These pairs can then be very efficiently associated on a Feshbach resonance \cite{Thalhammer2006} and subsequently transferred to the rovibronic ground state with STIRAP. The lattice has the advantage of shielding the molecules against inelastic collisions during the association process and subsequent state transfer. In particular, it should allow long STIRAP pulse durations, allowing us to resolve the weak hyperfine structure of ground state molecules \cite{Aldegunde2008}. As proposed by Jaksch {\em et al.} \cite{Jaksch2002}, dynamical melting of the lattice should ideally result in the formation of a BEC of molecules in the rovibronic ground state in a Mott-insulator-to-superfluid-type transition.

\section{Acknowledgements}
We are indebted to R. Grimm for generous support and we thank T. Bergeman, H. Salami, J. Hutson, J. Aldegunde, and E. Tiemann for valuable discussions. We gratefully acknowledge funding by the Austrian Ministry of Science and Research (BMWF) and the Austrian Science Fund (FWF) in form of a START prize grant and by the European Science Foundation (ESF) in the framework of the EuroQUAM collective research project QuDipMol. R.H. acknowledges support by the European Union in form of a Marie-Curie International Incoming Fellowship (IIF).

\clearpage
\begin{table}
\begin{center}
\caption{\label{tab1}Observed excited state levels in the wavelength range from 1118 nm to 1134 nm. Transitions were measured from Feshbach state $|s\!>$ to the first electronically excited state, addressing both $($A$^1\Sigma_u^+ - $b$^3\Pi_u) 0_u^+$ levels and $(1)^3\Sigma_g^+$ levels. Levels are given according to the excitation wavelength (WL) from $|s>$, which essentially corresponds to the $F\!=\!3,m_F\!=\!3$ two-atom asymptote. The data is taken at a magnetic field of 1.98 mT. Wavemeter accuracy is about 0.001 nm. The energy of these levels above the rovibronic ground state X$^1\Sigma_g^+$ $|v\!=\!0,J\!=\!0>$ is given in the second column, where the binding energy of the rovibronic ground state is taken from Ref.\cite{Danzl2008}. The assignment to either the coupled $($A$^1\Sigma_u^+ - $b$^3\Pi_u) 0_u^+$ system or to the $(1)^3\Sigma_g^+$ is based on the vibrational spacing and similarities in the substructure of the levels. The levels marked with $\ast$ have been used for dark resonance spectroscopy coupling to deeply bound levels of the X$^1\Sigma_g^+$ state \cite{Danzl2008}. The ability to couple to such levels unambiguously reflects an important singlet component stemming from the A$^1\Sigma_u^+$ state and therefore clearly assigns these levels to the $0_u^+$ system. The quantum numbers given for the $0_u^+$ levels are coupled channels quantum numbers derived from molecular structure calculations and bear an uncertainty of two in the absolute numbering. The calculations show that these levels have about 70\% A$^1\Sigma_u^+$ state contribution. Two further levels observed near 1120.17 nm and 1117.16 nm that belong to the $0_u^+$ progression are not given in the table since no further measurements have been done on these levels. The level near 1129.5 nm exhibits a somewhat richer structure than the other levels assigned to $0_u^+$ and than exemplified in Fig. 3. Levels assigned to the $(1)^3\Sigma_g^+$ state form a regular vibrational progression and show a more complex substructure than the levels attributed to the $0_u^+$ system, as exemplified in Fig. 6. For these levels, the transition wavelength to one of the most prominent features is given, since an in depth analysis of the rotational and hyperfine structure remains to be done. The vibrational numbering for the $(1)^3\Sigma_g^+$ levels is the same as in Ref \cite{Amiot1985}. }

\begin{tabular}{lll}\\
\\
\hline
WL [nm]  & \multicolumn{1}{l}{Energy above} & Assignment\\
&X$^1\Sigma_g^+$ $|v\!=\!0\!>$&\\
&  [cm$^{-1}$]&\\ \hline
1132.481 & 12458.875 & $0_u^+$ $|v'\!=\!221,J\!=\!1>$\\
1129.492 & 12482.245 & $0_u^+$\\
1126.173$\ast$ & 12508.332 & $0_u^+$ $|v'\!=\!225,J\!=\!1>$\\
1123.104$\ast$ & 12532.598 & $0_u^+$ $|v'\!=\!226,J\!=\!1>$\\

1133.680 & 12449.536 & $(1)^3\Sigma_g^+$ $|v'\!=\!32>$\\
1130.510 & 12474.274 & $(1)^3\Sigma_g^+$ $|v'\!=\!33>$\\
1127.379 & 12498.838 & $(1)^3\Sigma_g^+$ $|v'\!=\!34>$\\
1124.274 & 12523.334 & $(1)^3\Sigma_g^+$ $|v'\!=\!35>$\\
1121.196 & 12547.756 & $(1)^3\Sigma_g^+$ $|v'\!=\!36>$\\
1118.155 & 12572.013 & $(1)^3\Sigma_g^+$ $|v'\!=\!37>$\\

\hline
\end{tabular}
\end{center}
\end{table}

\clearpage

\begin{figure}[ht]
\begin{center}
 \includegraphics[width=12cm]{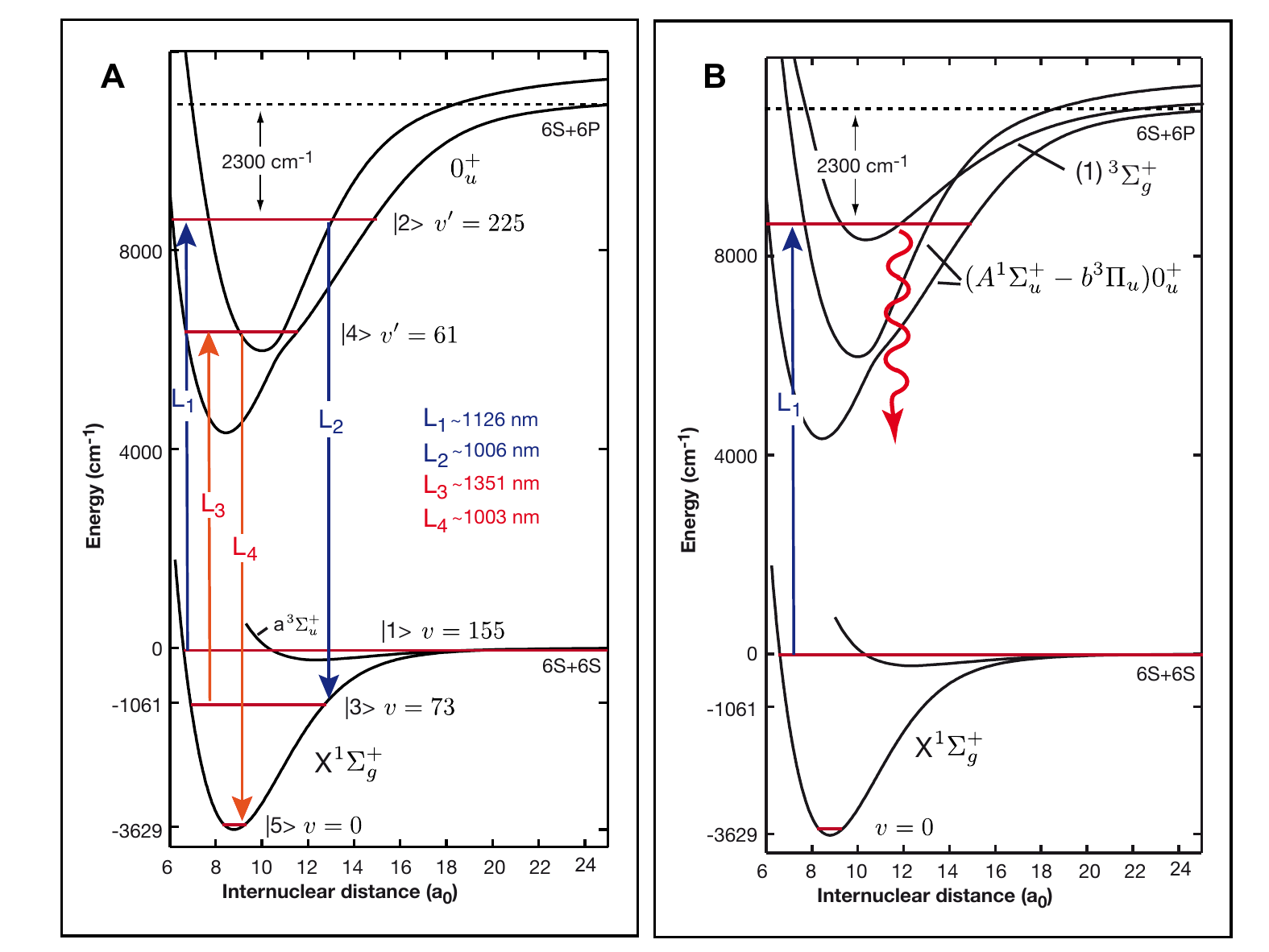}
 \caption{({\bf A}) Simplified molecular level scheme for Cs$_2$ showing the relevant ground state and excited state potentials involved in rovibrational ground state transfer. Molecules in a weakly bound Feshbach level $|1\!\!>= |v\!\approx\!155>$ (not resolved near the 6S$_\frac{1}{2}$ + 6S$_\frac{1}{2}$ two-atom asymptote, but shown in Fig.2) are to be transferred to the rovibrational ground state level $|5\!\!>=|v\!=\!0,J\!=\!0\!\!>$ of the singlet X$^1\Sigma_g^+$ potential with a binding energy of $3629$ cm$^{-1}$ by two sequential two-photon STIRAP processes involving lasers $L_1$ and $L_2$ near 1126 nm and 1006 nm and lasers $L_3$ and $L_4$ near 1351 nm and 1003 nm. The intermediate ground state level $|3\!\!>=|v\!=\!73,J\!=\!2\!\!>$ has a binding energy of $1061$ cm$^{-1}$.
({\bf B}) Probing candidate levels for $ |2\!\!> $ belonging to the electronically excited coupled $ ($A$^1\Sigma_u^+ - $b$^3\Pi_u) \ 0_u^+$ potentials. Here, we search for $|2\!\!>$ in loss spectroscopy with laser $L_1$ in a region near 8890 cm$^{-1}$ above the 6S$_\frac{1}{2}$ + 6S$_\frac{1}{2}$ asymptote, corresponding an excitation wavelength range of 1118 to 1134 nm. The wiggly arrow indicates loss from the excited levels due to spontaneous emission. Also shown is the excited $(1)^3\Sigma_g^+ $ potential, for which we find several levels.}
\end{center}
\label{fig1}
\end{figure}

\clearpage

\begin{figure}[ht]
\begin{center}
 \includegraphics[width=12cm]{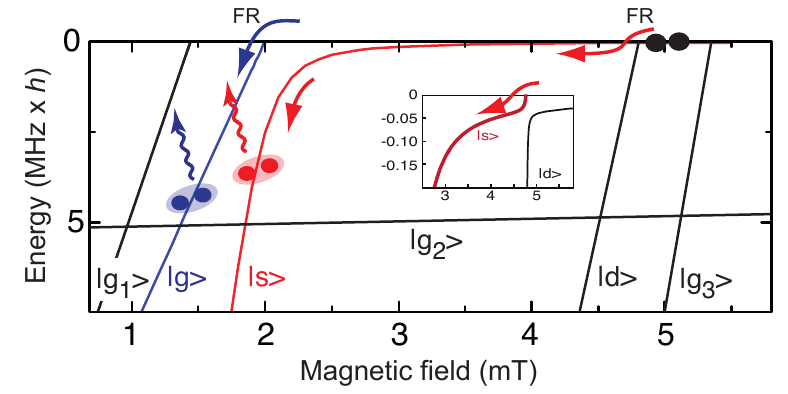}
 \caption{Initial Feshbach molecule production: Zeeman diagram showing the energy of weakly bound Feshbach levels \cite{Mark2007} and the Feshbach resonances (FR) used in the present work. The binding energy is given with respect to the $F\!=\!3, m_F\!=\!3$ two-atom asymptote. The molecules are produced either on a $d$-wave Feshbach resonance at 4.79 mT (see inset) and then transferred to the weakly bound $s$-wave state $|s\!\!>$ on an avoided state crossing, or on a $g$-wave Feshbach resonance at 1.98 mT, resulting in molecules in level $|g\!\!>$. In the first case, further lowering of the magnetic offset field to below $2.0$ mT changes the character of the $|s\!\!>$ level from open-channel to closed-channel dominated \cite{Mark2007}. The levels $|s\!\!>$ and $|g\!\!>$ are both candidate levels for the initial level $|1\!\!>$ shown in Fig.1. For completeness, further $g$-wave Feshbach levels, $|g_1\!\!>$, $|g_2\!\!>$, and $|g_3\!\!>$ are shown. Level $|g_2\!\!>$ connects $|g\!\!>$ to $|s\!\!>$ and can be used for Feshbach state transfer \cite{Mark2007}. Level $|g_3\!\!>$ is a further interesting candidate level for $|1\!\!>$ with low nuclear spin contribution \cite{Mark2007}.}
\end{center}
\label{fig2}
\end{figure}

\clearpage

\begin{figure}[ht]
\begin{center}
 \includegraphics[width=\textwidth]{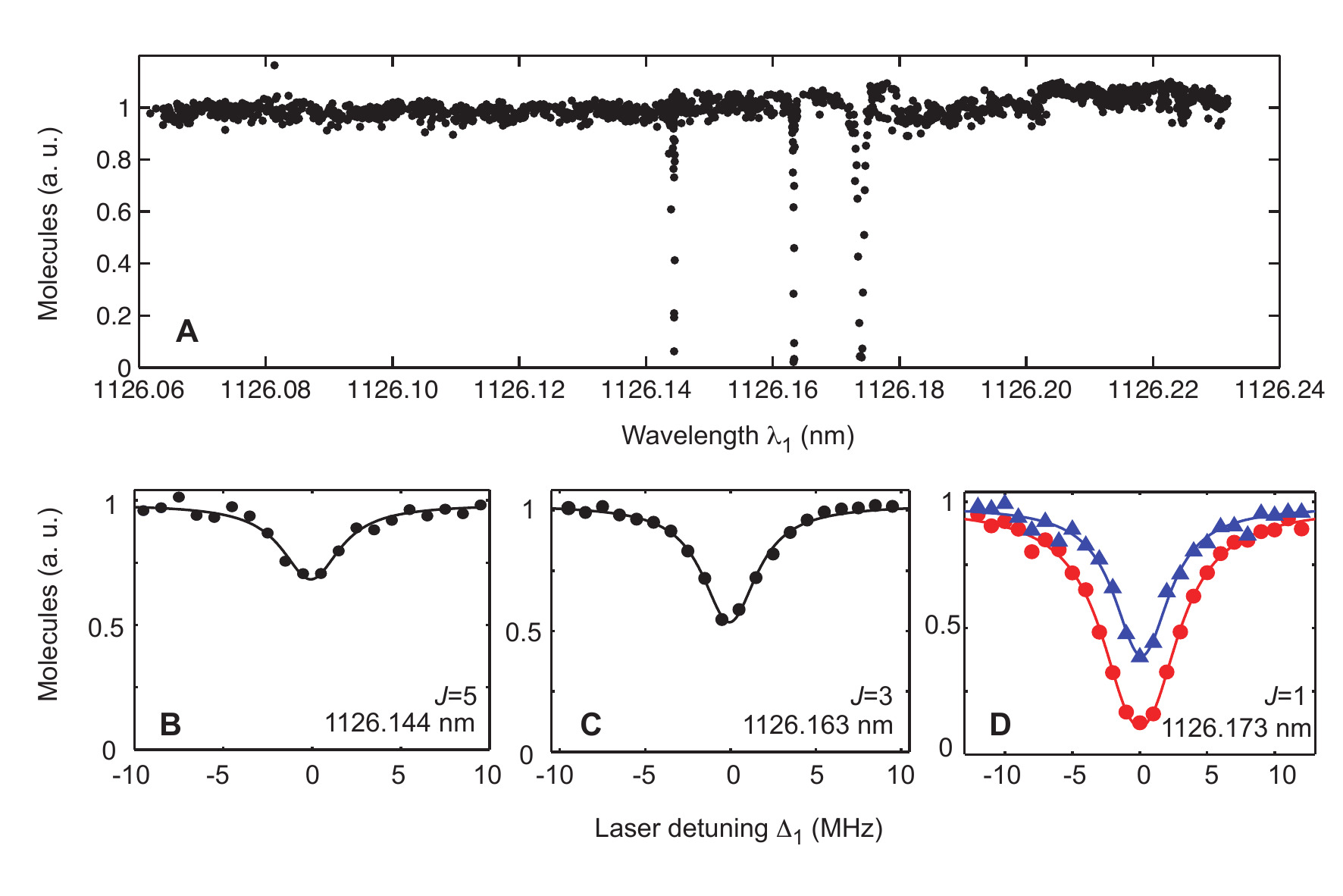}
 \caption{Loss resonances for excitation from the initial Feshbach level $|s\!\!>$ to the $0_u^+$ system. (\textbf{A}) Typical scan showing the number of molecules in $|s\!\!>$ as a function of laser wavelength $\lambda_1$ near 1126 nm. Three resonances can be identified, corresponding to $|J\!=\!5\!\!>$, $|J\!=\!3\!\!>$, and $|J\!=\!1\!\!>$, from left to right. The sample is irradiated with laser light at an intensity of $1\times 10^{6}$ mW/cm$^2$ for $\tau= 200$ $\mu$s. The laser is locked to a narrow band optical resonator that is tuned via a piezoelectric element with a step size of approximately 40 MHz. Wavelength is measured on a home-built wavemeter. The molecule number is normalized to the atom number measured in the same individual realization of the experiment to cancel out fluctuations that stem from shot-to-shot atom number fluctuations and the baseline is set to 1. (\textbf{B}), (\textbf{C}), and (\textbf{D}) represent measurements of the three individual lines with $|J\!=\!5\!\!>$, $|J\!=\!3\!\!>$, and $|J\!=\!1\!\!>$ at reduced intensity in order to avoid saturation. The solid lines represent fits as described in the text. The spectroscopy laser is stabilized to an optical resonator and the resonator is in turn referenced to an optical frequency comb, which allows precise and reproducible tuning of the frequency. The transition to $|J\!=\!1>$ in panel (\textbf{D}) is recorded at an intensity of $1.5 \times 10^4$ mW/cm$^2$ (circles) and $6 \times 10^3$ mW/cm$^2$ (triangles), (\textbf{B}) and (\textbf{C}) are recorded at $1 \times 10^6$ mW/cm$^2$ and $2 \times 10^5$ mW/cm$^2$, respectively. Pulse duration is $\tau=10$ $\mu$s. }
\end{center}
\label{fig3}
\end{figure}

\clearpage

\begin{figure}[ht]
\begin{center}
 \includegraphics[width=10cm]{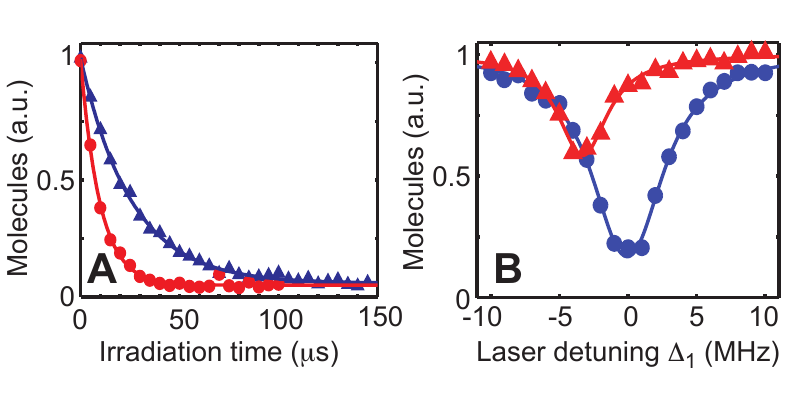}
 \caption{Loss of molecules for excitation near 1126.173 nm from Feshbach level $|s\!\!>$.
 (\textbf{A}) Time dependence of molecular loss on resonance at 1126.173 nm for two different laser intensities, $5.7 \times 10^5$ mW/cm$^2$ (circles) and $2.1 \times 10^5$ mW/cm$^2$ (triangles). The magnetic offset field is 1.9 mT. The fitted exponential decay gives the decay constants $\tau = 9.7\pm0.6\ \mu$s (circles) and $\tau = 25.5\pm1\ \mu$s (triangles).
 (\textbf{B}) Loss of molecules in $|s\!\!>$ as a function of laser detuning $\Delta_1$ near 1126 nm with an irradiation time of $\tau=10\,\mu$s for two values of the magnetic field, $1.9$ mT (dots) and $2.2$ mT (triangles). In both cases, the excited state spontaneous decay rate was determined to $\approx 2 \pi \times 2\,$MHz. At higher magnetic fields, Feshbach level $|s\!\!>$ acquires more open-channel character, reducing radial wave function overlap with the excited rovibrational levels. The shift in transition frequency is the result of a differential magnetic field shift of the Feshbach level $|s\!\!>$ and the excited state level.}
\end{center}
\label{fig4}
\end{figure}

\clearpage

\begin{figure}[ht]
\begin{center}
 \includegraphics[width=12cm]{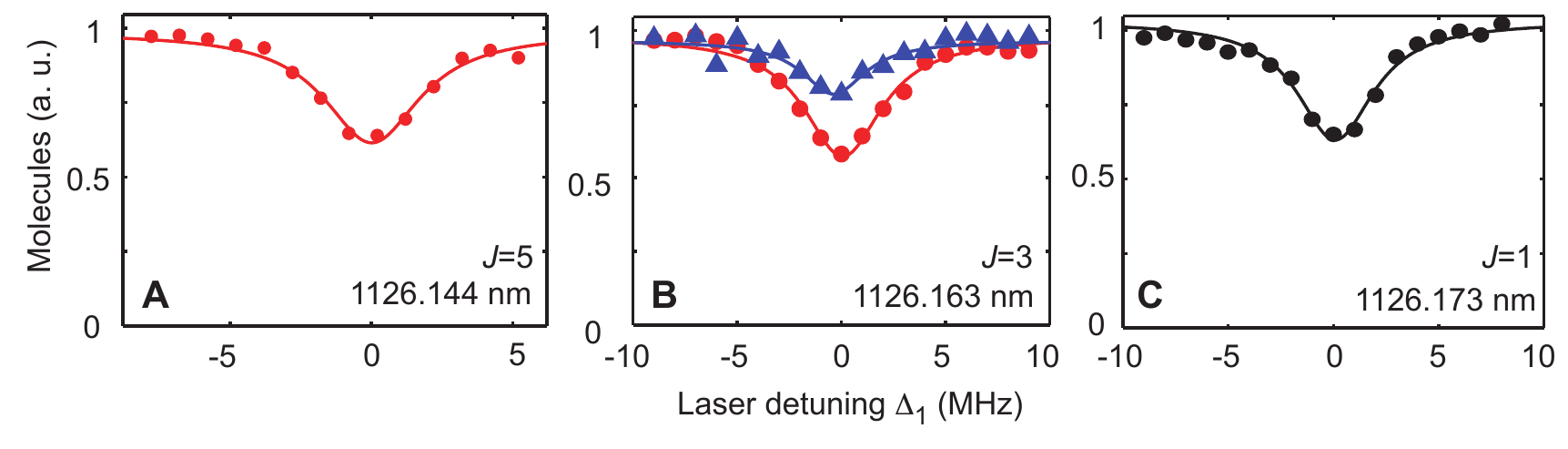}
 \caption{Loss resonances for excitation from the initial Feshbach level $|g\!\!>$. \textbf{(A)},\textbf{(B)}, and \textbf{(C)} show the loss for excitation to $|J\!=\!5\!\!>$, $|J\!=\!3\!\!>$, and $|J\!=\!1\!\!>$, corresponding to the resonances shown in Fig.3. The laser intensities are $1.5 \times 10^4$ mW/cm$^2$ for panel \textbf{(A)} and for the circles in panel \textbf{(B)}. The second resonance in \textbf{(B)} (triangles) is measured with $5.6 \times 10^3$ mW/cm$^2$. \textbf{(C)} The line at 1126.173 nm is measured at $1 \times 10^6$ mW/cm$^2$. All measurements are done with an irradiation time of $\tau=10$ $\mu$s. From a series of such measurements at different intensities we determine the line strengths for $|J\!=\!5\!\!>$, $|J\!=\!3\!\!>$, and $|J\!=\!1\!\!>$ to  $\Omega_1\!=\!2\pi\!\times\!1$ kHz $ \sqrt{I/(\mathrm{mW/cm}^2)}$,  $\Omega_1\!=\!2\pi\!\times\!1$ kHz $ \sqrt{I/(\mathrm{mW/cm}^2)}$, and $\Omega_1\!=\!2\pi\!\times\!0.1$ kHz $ \sqrt{I/(\mathrm{mW/cm}^2)}$, respectively.}
\end{center}
\label{fig5}
\end{figure}

\clearpage

\begin{figure}[ht]
\begin{center}
 \includegraphics[width=13cm]{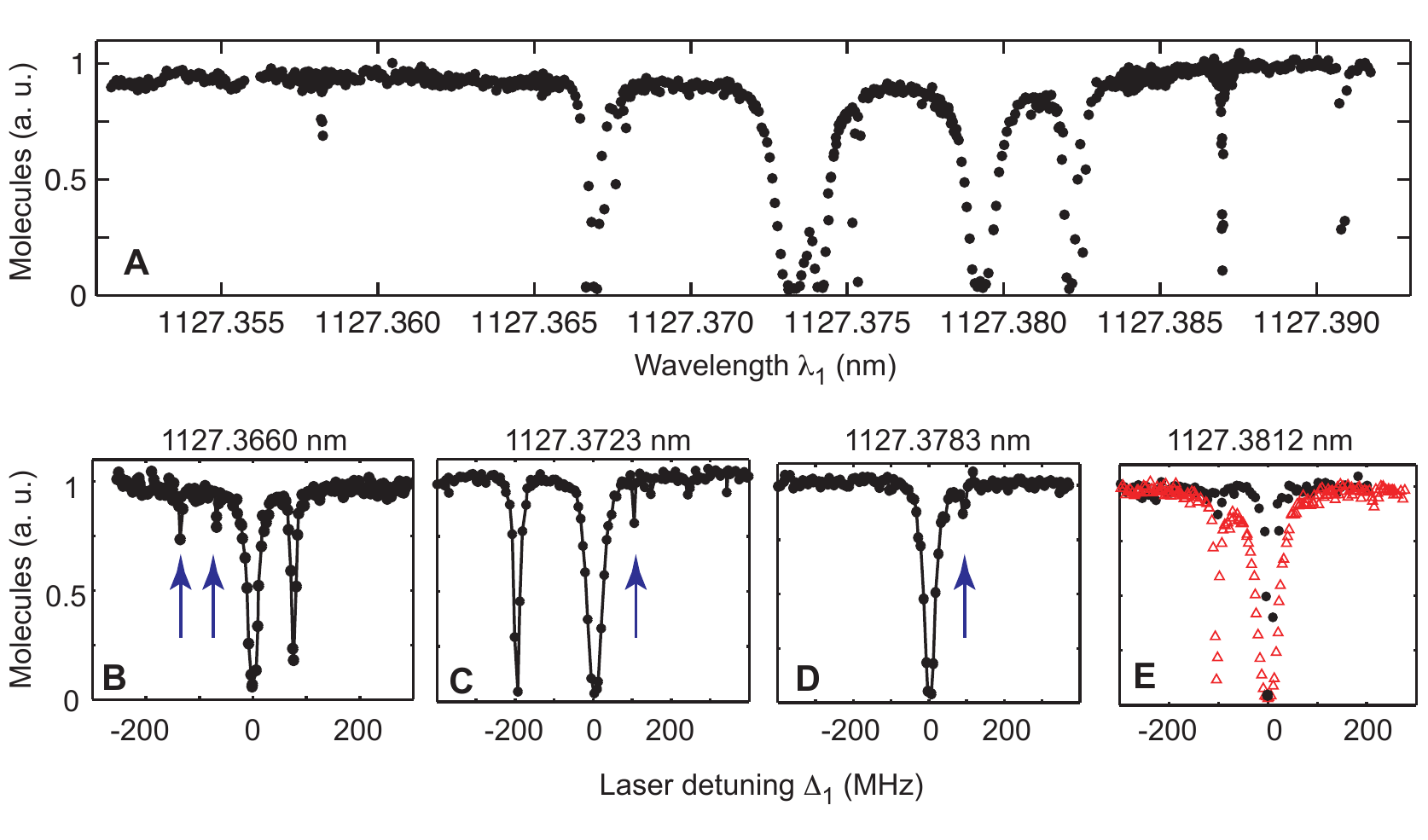}
 \caption{Loss of molecules for excitation near 1127.17 nm from Feshbach level $|s\!\!>$ to the triplet $(1)^3\Sigma_g^+$ state.
 (\textbf{A}) represents a broad scan with laser irradiation at an intensity of $5 \times 10^5$ mW/cm$^2$ for $\tau= 100$ $\mu$s at a step size of 20 MHz. A rich structure due to rotation and excited state hyperfine splitting can be seen which is qualitatively different from the spectrum shown in Fig.3. The lines are greatly broadened by the high intensity and long irradiation time. The spectroscopy laser is locked to a narrow band optical resonator that is stepped via a piezoelectric element. Scans of about 750 MHz were recorded as a function of piezo voltage on the resonator. Voltage was converted to wavelength for each scan by a linear interpolation. (\textbf{B})-(\textbf{E}) represent scans over some of the observed features at a reduced intensity of $8 \times 10^4$ mW/cm$^2$ and an irradiation time of $\tau= 10$ $\mu$s in order to reduce broadening of the lines. The step size is about 7 MHz. Resonator piezo voltage is converted to frequency with an estimated error of 10 \%. The vertical arrows indicate weak lines that have been verified in additional scans with higher power. In panel  (\textbf{E}) the power was somewhat increased for an additional measurement (triangles) that emphasizes such a weak line. The wavelengths given to identify the zero point on the frequency axis for each subpanel are not meant to imply this level of accuracy which is limited to 0.001 nm by wavemeter calibration. Nevertheless, they give a measure of the energy of the sublines relative to each other.}
\end{center}
\label{fig6}
\end{figure}

\end{document}